\documentclass[fleqn,twoside]{article}
\usepackage{espcrc2}
\usepackage{xspace}
\usepackage{graphicx}

\newcommand{\superk}{Super-Kamkiokande\xspace}

\newcommand{\dms}       {$\Delta m^2$\xspace}
\newcommand{\sstt}      {$\sin^2 2 \theta$\xspace}

\title{Recent Results from the K2K Experiment}

\author{Christopher W. Walter for the K2K Collaboration \\
  590 Commonwealth Ave. Physics Dept. Boston University, Boston MA,
  02215 USA.\thanks{email address: walter@budoe.bu.edu}}
       
\begin{document}

\begin{abstract}
  The K2K experiment has collected approximately half of its
  allocated protons on target between June of 1999 and July of 2001.
  These proceedings give a short introduction to the experiment and
  summarize some of the recent results.  
\vspace{1pc}
\end{abstract}

\maketitle

\section{INTRODUCTION}

The K2K experiment uses an accelerator produced neutrino beam with a
long baseline between the production of the neutrinos and their
detection to search for neutrino oscillations.  The goal of the K2K
experiment is to confirm the atmospheric neutrino oscillation effect
by observing neutrino oscillation over the 250~km distance between KEK
and Super-Kamiokande.

\section{EXPERIMENTAL DETAILS}

The neutrino beam used in the K2K experiment is produced by a 12~GeV proton
beam taken from the KEK Proton Synchrotron(PS) with fast extraction.  After
hitting an aluminum target the positively charged particles, mostly pions,
are focused by a pair of horns. The beam is approximately 98\% muon neutrino
with a 2\% electron neutrino contamination. The peak energy of the resulting
neutrinos is 1~GeV, with a mean energy of 1.4~GeV. The beam pulse is
extracted from the PS in a single turn every 2.2 seconds with a pulse
structure of 9 bunches in 1.1~$\mu$sec.

\begin{figure}[htb]
  \begin{center}
    \includegraphics[width=2.5in]{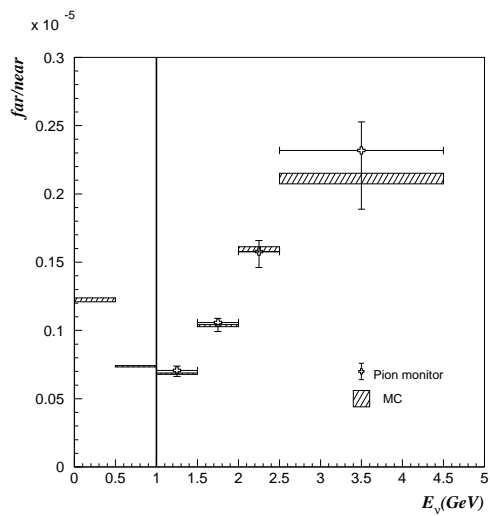}
    \caption{Ratio of the neutrino spectrum at Super-K and the near
      detectors as predicted by the beam MC and measured by the
      pion monitor for neutrino energies above 1~GeV.}
    \label{fig:far-near}
  \end{center}
\end{figure}

The pion beam's momentum and divergence is occasionally monitored by a
gas Cherenkov monitor downstream of the second horn. These
measurements are used to confirm the correctness of the beam MC. The
beam MC is used to predict the neutrino flux at Super-K given the
measurements at the near detector at the point of neutrino production.
Figure~\ref{fig:far-near} shows both the measured and predicted ratio
of neutrino fluxes at the far and near detectors using the pion
monitor.

\begin{figure}[htbp]
  \begin{center}
    \includegraphics[width=3.0in]{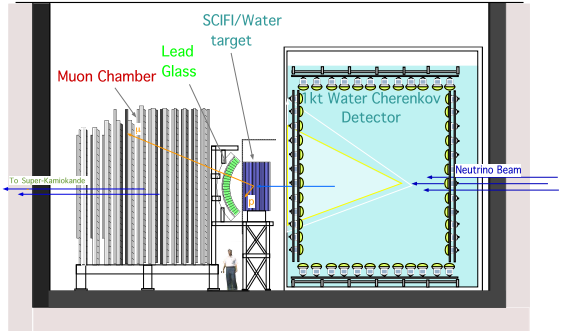}
    \caption{The near detector of the K2K experiment.  The detector is 
      compromised of a 1~kt water Cherenkov detector, a
      scintillating-fiber water-target calorimeter, a lead-glass
      calorimeter, and a set of muon rangers.  The experimental hall
      is cylindrical with a 24~m diameter and a depth of 16~m.}
    \label{fig:k2k-nearDetector}
  \end{center}
\end{figure}

The near detector of the K2K experiment is 300~m downstream of the beam
target. The near detector is comprised of two detector systems: a 1~kt water
Cherenkov tank and a fine-grained tracking system.
Figure~\ref{fig:k2k-nearDetector} is a schematic representation of the near
detector. The 1~kt water Cherenkov detector is a scaled down version of the
Super-Kamiokande detector. Its main purpose is to use the same method of
detecting events at both the near and far detector sites. In this way,
systematic errors or biases introduced by using a water Cherenkov detector
will be canceled in a direct comparison of measurements in the near and far
detectors.  It is also desirable to make precision measurements of the
transverse profile, energy distribution, and $\nu_e$ contamination of the
neutrino beam at the near detector.  For this reason, a fine-grained detector
is also employed.  We also hope to measure exclusive neutrino reactions on
water such as single-$\pi$ production. The fine-grained detector consists of
a scintillating fiber tracker, trigger counters, lead glass counters and a
muon ranger.

The lead glass counters are used to determine the $\nu_e$
contamination in the $\nu_\mu$ beam.  Electrons produced by electron
neutrinos shower in the lead glass and their energy can be measured
with a resolution of 8\%/$\sqrt{E_{e}}$.  Muons, on the other hand,
are minimum ionizing and leave less than $\sim$ 1~GeV.  The reason it
is important to measure accurately the amount of contamination by
electron neutrinos is that unmeasured $\nu_e$s in the neutrino beam
might be interpreted at the far detector as evidence for $\nu_\mu
\rightarrow \nu_e$ oscillation.  A preliminary analysis has shown a
measured electron contamination consistent with the expectation of
$\approx$~2\%.  

\begin{figure}[!htbp]
  \begin{center}
    \includegraphics[width=3.1in]{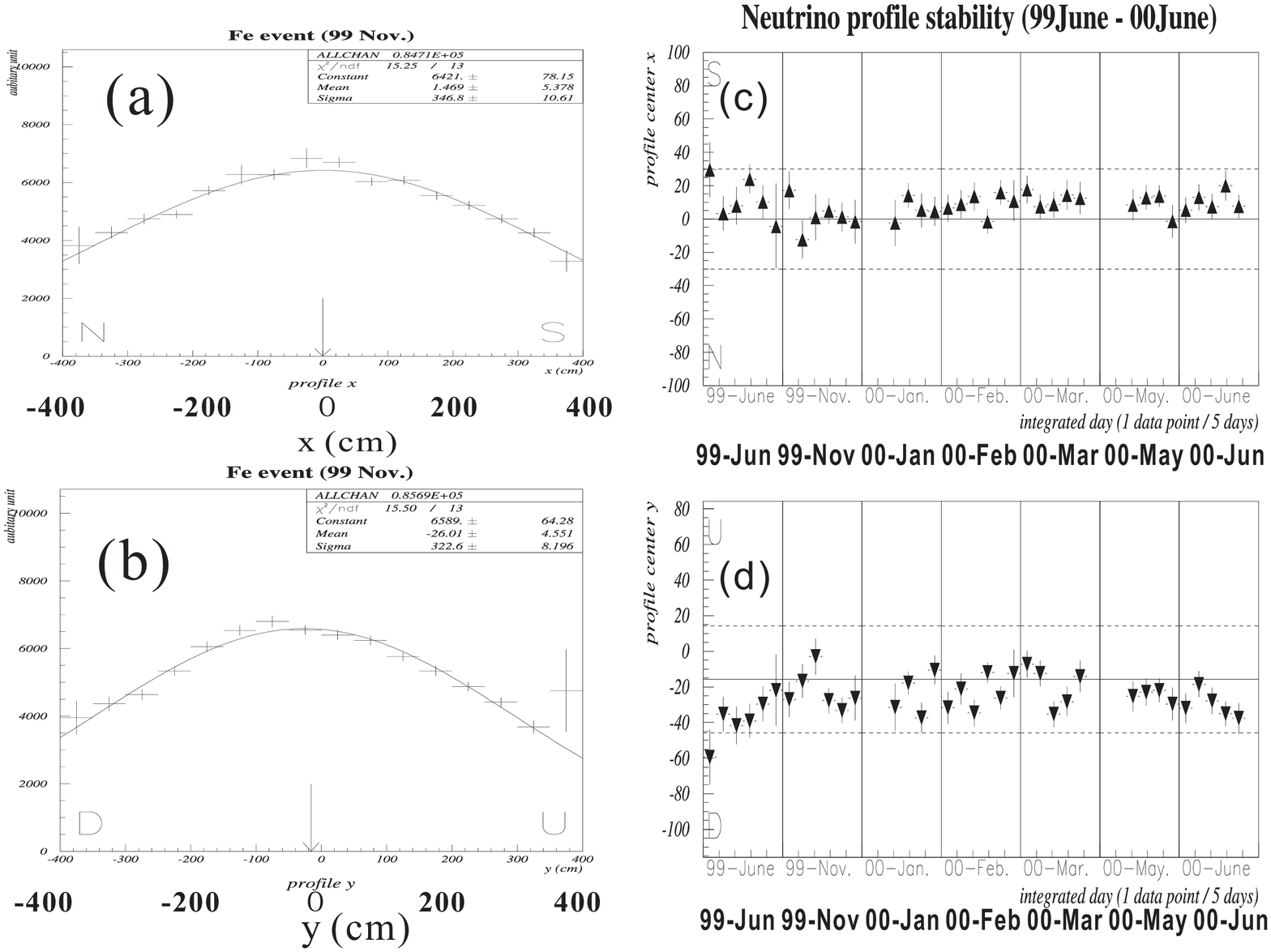}
    \caption{Horizontal and vertical beam profiles as measured by muon
      chambers.  Also shown are the peak positions as a function of
      time.  The position is plotted on a time scale of every five
      days.  The solid line shows the direction to Super-K and the
      dashed lines show a $\pm$1~mrad deviation.}
    \label{fig:muc-profile}
  \end{center}
\end{figure}

Finally, the muon rangers are made of 12 iron plates instrumented with
drift tubes. They are used to measure the momentum of muons generated
by charged current interactions in the water target of the fiber
tracker. Also because of the large mass of the iron there is a very
large contained event rate which allows us to measure the neutrino
beam center and profile as a function of time.
Figure~\ref{fig:muc-profile} shows the horizontal and vertical beam
profiles as measured in the muon chambers and also the peak position
as measured every five days.  This measurement confirms the the beam
direction is stable to within 1~mrad.

\subsection{CURRENT STATUS AND RESULTS}

By comparing the GPS time stamps of proton beam spills at KEK and Super-K
triggers we search for events that have arrived at Super-K within 1.1~usec of
the the beginning of the neutrino pulse.  Two classes of events are
reconstructed at Super-K.  The first, known as ``Fully-Contained(FC)'' are
events where all of the Cherenkov light in the event is contained in the
inner detector of Super-K.  The second class of events are known as ``outer
detector(OD)'' events since Cherenkov light is also seen in the outer
detector.

The FC events are further sub-divided into events which fall inside of
the 22.5~kton fiducial volume(FC-in) and those that fall
outside(FC-out).  Since the Super-K reconstruction algorithms are, at
this time, only guaranteed to work inside of the fiducial volume we
only use the FC-in events for quantitative analysis.

By using the measured neutrino rates in the near detectors and the
ratio of near to far fluxes as calculated by the beam MC we predict
how many events should be seen at Super-K.  The number predicted in
the fiducial volume of Super-K by the near detectors for our present
running period is: $N_{SK}^{pred}\ = 80.1^{+6.2}_{-5.4}$, compared to
56 events detected.  The number of observed events are summarized in
Table~\ref{tbl:SKobs}.

\begin{table}[htb]
  \centering
  \caption{Observed number of events in the fiducial volume at SK.}
  \vspace{.6pc}
  \begin{tabular}{ll|rl}
    \hline\hline
    single ring  & $\mu$-like        &   30 &       \\
                 & e-like            &    2 &      \\
    \multicolumn{2}{c|}{multi  ring} &   24 &      \\
    \multicolumn{2}{c|}{Total}       &   56 &      \\
    \hline\hline
  \end{tabular}
  \label{tbl:SKobs}
\end{table}

\begin{figure}[!htb]
  \begin{center}
    \includegraphics[width=3.1in]{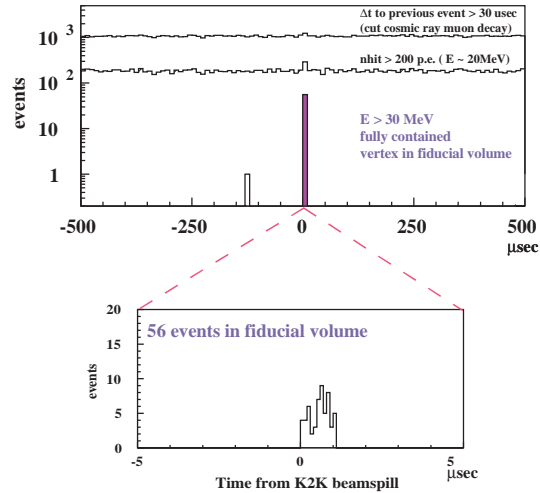}
    \caption{Beam Spill time - SK time - TOF for FC events in
      Super-K.  The upper figure shows a simplified view of the K2K
      data reduction. The lower figure shows the events detected in
      the fiducial region of Super-K.  All events fall in the beam
      window.}
    \label{fig:tdiff}
  \end{center}
\end{figure}

Figure~\ref{fig:tdiff} shows the time difference of the FC events
between the KEK accelerator beam time and the time of the trigger in
Super-K corrected for the time of flight of the neutrinos.  The FC
events in the fiducial volume are shown in the lower figure and all of
the events clearly fall within the 1.1~$\mu$sec beam window.

\begin{figure}[!htb]
  \begin{center}
    \includegraphics[width=2.4in, angle=-90]{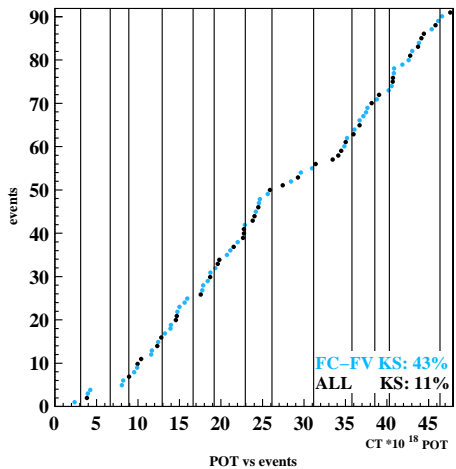}
    \caption{Arrival time of each far-detector event versus
      accumulated protons-on-target. Light blue circles represent the
      cleanest events: fully-contained with vertex in the fiducial
      volume. The dark blue circles includes out-of-fiducial-volume
      events.}
    \label{fig:ctgap}
  \end{center}
\end{figure}

Although fully-contained events will provide the cleanest sample of
beam-induced neutrinos, we are also looking for neutrinos that
interact in the Super-K outer detector, or in the rock outside the
detector.  This has the potential to increase the number of events
usable in the oscillation analysis. More importantly, it provides an
important cross check of beam stability during periods when a low
statistics fluctuation suggests a gap in the data; evidence of OD
events indicates that any gap in the golden inner detector event
sample is statistical in nature. In Fig.~\ref{fig:ctgap} the arrival
time of Super-K events is plotted versus integrated luminosity.  The
arrival times of the events are consistent with a Poisson distributed
distribution.  

In addition to a reduction in the number of events seen, in the case
of oscillations we expect to see a spectral distortion.  We can
estimate the neutrino energy for single-ring events from quasi-elastic
scattering. Since the direction of the incoming neutrino is known, by
measuring the momentum and scattering angle of the single muon ring,
the energy of the neutrino can be calculated.

\section{OSCILLATION ANALYSIS}
\label{sec:oscillation-analysis}

The oscillation analysis is performed by the maximum-likelihood
method.  The major features of the analysis are that both the spectral
shape information at SK and the number of observed events at SK are
used.  Also, correlations between energy bins in the neutrino spectrum
along with correlation in the far/near ratio are taken into account.

The criteria used to select the neutrino events in SK have been
described in detail elsewhere~\cite{Ahn:2001cq}. The basic selection
criteria are that events must have no activity in the outer detector,
have an electron equivalent energy greater than 30 MeV, and have a
vertex reconstructed inside of the 22.5~kton fiducial volume.  In
order to predict the number of events and neutrino spectrum observed
in SK, the number of events observed in the 1kt detector and the
spectrum measured in all of the near detectors were extrapolated to
250~km away by using the far/near spectrum ratio estimated by the MC
and the pion monitor measurements. In this way, the systematic error
on the the predicted number of events is greatly reduced thanks to the
similar efficiencies of the 1kt and SK and the fact that they both
contain a water target.

\subsection{Description of the analysis techniques}

In order to extract the maximum amount of information about the mixing
parameters from the 56 events detected in the fiducial volume of
\superk we performed an unbinned maximum likelihood analysis, comparing
the data with MC expectation. For the single-ring events in \superk,
where the energy of the incoming neutrino can be easily estimated,
both the shape of the energy spectrum and the total number of events
which were detected were considered in the likelihood function.  For
the all other events, where QE kinematics can't be used to estimate
the neutrino energy, only the total number of detected and expected
events are compared.

The likelihood is composed of the product of these two observables.
${\cal L}(\Delta m^2, \sin^22\theta) = {\cal L}_{norm}(\Delta m^2,
\sin^22\theta) \times {\cal L}_{shape}(\Delta m^2, \sin^22\theta)$.
The normalization term is just the Poisson probability to observe
$N_{obs}$ when expected number of events is $N_{exp}(\Delta m^2,
\sin^22\theta)$.  The shape term is the product of the probabilities
that each event will have the measured reconstructed neutrino energy
if it was drawn from the energy spectra as estimated by our MC
simulation.  These two terms are correlated since a distortion in the
spectrum results in a decrease in the total number of events.  In
addition, the MC expectation is a function of a set of systematic error
parameters.  Uncertainties on the measured neutrino flux, far/near
ratio etc will effect the expectation at Super-K.

The basic form of the likelihood is then:

\begin{equation}
  \label{eq:basic-likelihood}
  {\cal L} = 
  { \mu^{N_{obs}} e^{-\mu} \over N_{obs} ! }
  \prod_{i=1}^{N_{1R}} p_{i}(E_i, \Delta m^2, \sin^2 2 \theta) ,
\end{equation}

\noindent
where $\mu$ is the expected rate of total observed events at this \dms and
\sstt, $N_{obs}$ is the number of total events actually observed,
$N_{1R}$ is the number of single-ring events, and $p_i$ is the
probability of the $i^{th}$ event having the energy $E_i$ at a
particular set of oscillation parameters.  

In order to make allowed regions we calculated $\cal{L}$ over the
space of oscillation parameters \dms and \sstt.  In order to
incorporate the systematic errors into this procedure two
complimentary techniques were used.  The first method uses numerical
propagation of the systematic errors. The error matrices that were
determined for the near detector flux, the near far extrapolation, and
the shape and normalization as measured at \superk were used to
generate a set of Gaussian random correlated systematic error numbers
which were then applied to our MC expectation.  For example, the
measured neutrino flux was sampled within its fitted errors which
changes the expectation for the measured spectral shape at Super-K.
This procedure was repeated 500 times at each point in the oscillation
space and values of the likelihood were averaged so that the resulting
likelihood was the likelihood sampled properly over the systematic
error parameters.  In this way the likelihood was sampled properly
with all the correlations between the systematic error parameters
taken into account.

The other technique used to include systematic errors was to modify
the likelihood by adding constraint terms associated with each set of
systematic error parameters. In this case the form of the likelihood
becomes:

\begin{equation}
  \label{eq:error-likelihood}
  {\cal L}_{sys} = {\cal L} \times \prod_{i=0}^{N_{par}} 
  e^
  {\left(\Delta{\bf f}^T_i \cdot {\bf M}^{-1}_i \cdot \Delta{\bf f}_i
    \right)} ,
\end{equation}

\noindent
where {$\cal L$} is from Eqn~\ref{eq:basic-likelihood}, $N_{par}$ is
the number of systematic error matrices, ${\bf M}$ is each error
matrix, and $\Delta {\bf f}_i = {\bf f}_i - {\bf f}_i(default)$ is the
vector of systematic parameters associated with that matrix where
${\bf f}(default)$ is the default value of the parameter without any
modification.  At each point in the oscillation parameter space the
systematic error parameters are varied in the constraints and in the
main part of the likelihood until the total likelihood is minimized.

The set of systematic errors which were considered for this analysis
included the spectrum measurement at the near detectors, the non-QE/QE
cross-section ratio, and the far/near flux ratio.  These errors are
dependent on energy and are represented by a set of matrices with the
full energy correlations included.  In addition there are errors which
represent the fiducial volume error in the 1kt and SK, the uncertainty
on the energy scale of SK, and the uncertainty in the detection
efficiency of SK.

The target radius and horn current, and hence, the neutrino spectrum
in Jun 99 were different from those used in the rest of the running
period.  The full analyses of the near detector spectra and far/near
ratio including all correlations has not yet been completed for this
data period. Therefore, for ${\cal L}_{shape}$, the events in Jun 99 are
not considered.  For this reason 29 1-ring $\mu$-like events are used
in the likelihood.  For $L_{norm}$, the data of whole experimental
period are used, i.e. $N_{obs}= 56$.

\subsection{Fit results}

The ${\cal L}$ was calculated by either numerically propagating the
systematic errors, or adding the likelihood constraint terms and
minimizing the likelihood, at each point in the \dms and \sstt space.
Then, the point where the likelihood was maximized was located.  The
resulting best fit oscillation parameters are summarized in Table
\ref{tbl:best}.

\begin{table*}[htb]
  \caption{Best fit points. The unit of $\Delta m^2$ is
    $10^{-3}$~eV$^2$.  The results are shown for both the entire
    parameter space and for the case where the parameters are
    constrained to be physical.}
  \renewcommand{\tabcolsep}{2pc} 
  \renewcommand{\arraystretch}{1.2} 
  \begin{tabular}{l|lc|lc}
    \hline\hline
    & \multicolumn{2}{|c|}{Likelihood}& \multicolumn{2}{c}{Numerical}\\
    & \multicolumn{2}{|c|}{Constraint}&  \multicolumn{2}{c}{Propagation}\\
    & \sstt & \dms& \sstt & \dms \\
    \hline
    Shape only          & 1.0  & 3.0 &  1.0  & 3.2 \\
    (allowing unphysical)  & 1.09 & 3.0 &  1.05 & 3.2 \\
    \hline
    Norm + Shape        & 1.0  & 2.8 &  1.0  & 2.7 \\
    (allowing unphysical)  & 1.03 & 2.8 &  1.05 & 2.7 \\
    \hline\hline
  \end{tabular}

  \label{tbl:best}
\end{table*}

\begin{figure}[!htb]
  \begin{center}
    \includegraphics[width=3.0in]{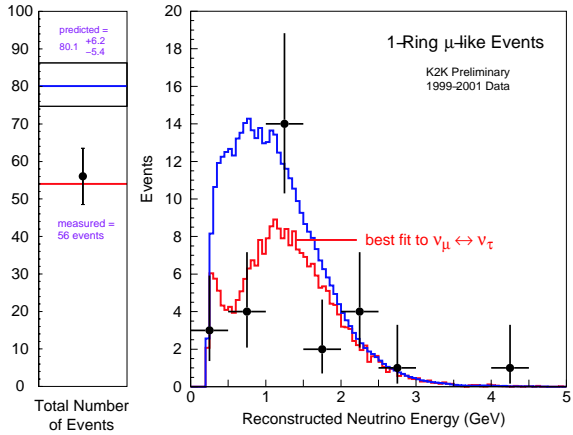}
    \caption{The total event count and single-ring $\mu$-like
      neutrino spectrum overlaid with
      expectations for no oscillations and oscillations at the
      K2K best fit point.}
    \label{fig:k2k-nuosc-result}
  \end{center}
\end{figure}

At the best fit point the total number of predicted events is 54.2(to
be compared with the 56 observed).  Using the single-ring $\mu$-like
events as a fairly pure sample of quasi-elastic interactions, we
measure the neutrino energy spectrum.  The distribution of data is
very low in second bin, from 500 MeV to 1 GeV. This is consistent with
neutrino oscillations, as shown by the overlay of the best-fit
spectrum in Fig.~\ref{fig:k2k-nuosc-result}.  The best-fit parameters
are $\Delta m^2 = 2.7 \times 10^{-3} {\rm eV}^2$ and $\sin^2 2\theta =
1$.

The consistency between the observed and best-fit reconstructed
$E_\nu$ spectrum shown in Fig~\ref{fig:k2k-nuosc-result} was checked
by the use of the KS test. A KS probability of 79\% was obtained.
Both our best fit number and shape agree very well with the
observations.

\subsection{The probability of the no oscillation hypothesis}

\begin{figure}[!htb]
  \begin{center}
    \includegraphics[width=3.0in]{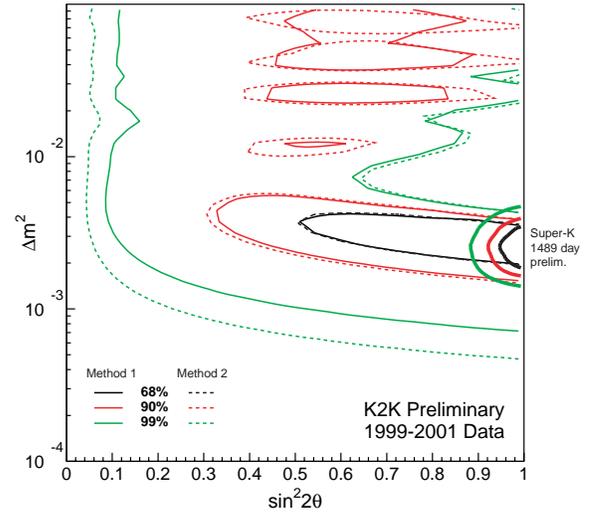}
    \caption{The confidence intervals based on the combined
      analysis of the total event rate plus the neutrino spectrum.
      The two methods refer to different treatments of systematic
      uncertainties.}
    \label{fig:results-of-best-fit}
  \end{center}
\end{figure}

The probability that our result is due to a statistical fluctuation
instead of neutrino oscillation is calculated by computing the
likelihood ratio of the best fit point to the no oscillation case. The
results are 0.7\% using the likelihood constraint method, and 0.4\%
for the method using the numerical propagation of errors.  Finally,
allowed regions of oscillation parameters for both methods are drawn
in Fig.~\ref{fig:results-of-best-fit} along with the Super-K allowed
region.  The allowed regions from the two methods are consistent with
each other as are the no-oscillation probabilities from the two
methods.  

Both methods give essentially the same results, the 90\% CL
contour crosses with the $\sin^22\theta=1$ axis at $1.5$ and
$3.9\times 10^{-3}$~eV$^2$.  The results are clearly in agreement with
the parameters found by the Super-Kamiokande atmospheric neutrino
analysis. The probability that our data is a fluctuation from the
no-oscillation hypothesis is less than 1\%.

The K2K experiment has collected approximately one-half of its
expected protons-on-target.  We expect to begin new running in January
of 2003.


\end{document}